\documentstyle[12pt,report,epsfig]{article}

\def\be{\begin{equation}}
\def\ee{\end{equation}}
\def\bea{\begin{eqnarray}}
\def\eea{\end{eqnarray}}

\def\Osz{{\mbox{$\bar{\nu}_{\mu}$}}$\rightarrow${\mbox{$\bar{\nu}_{e}$}}}
\def\el{\mbox{e$^-$}}
\def\pos{\mbox{e$^+$}}
\def\nue{\mbox{$\nu_e$}}
\def\nueb{\mbox{$\bar{\nu}_{e}$}}
\def\num{\mbox{$\nu_{\mu}$}}
\def\numb{\mbox{$\bar{\nu}_{\mu}$}}
\def\nux{\mbox{$\nu_{x}$}}
\def\mum{\mbox{$\mu^-$}}
\def\mup{\mbox{$\mu^+$}}
\def\pim{\mbox{$\pi^-$}}
\def\pip{\mbox{$\pi^+$}}
\def\Nzg{\mbox{$^{12}$N$_{\rm g.s.}$}}
\def\Cng{\mbox{$^{12}$C\,(\,\nue\,,\,\el\,)\,\Nzg}}

\def\Png{\mbox{ p\,(\,\nueb\,,\,e$^{+}$\,)\,n}}
\def\dm{\mbox{$\Delta$m$^2$}}
\def\sn{\mbox{sin$^2\,(2\,\Theta)$}}

\def\Gdng{\mbox{Gd\,(\,n,\,$\gamma$\,)}}
\def\Gn{\mbox{Gd\,(\,n,\,$\gamma$\,)}}
\def\Pn{\mbox{p\,(\,n,\,$\gamma$\,)}}
\def\eVc{\mbox{eV$^2$/c$^4$}}

\def\lmc{\mbox{$\Delta log[L(\Delta\rm{m}^2;\rm{sin}^2\,2\Theta)]_{MC}$}}
\def\lex{\mbox{$\Delta log[L(\Delta\rm{m}^2;\rm{sin}^2\,2\Theta)]_{EXP}$}}

\begin{document}
\renewcommand{\floatpagefraction}{0.95}

\thispagestyle{empty}

\vspace*{4cm}

\title{The Search for Neutrino Oscillations \Osz\ with KARMEN}

\author{ T.E. JANNAKOS }

\address{Forschungszentrum Karlsruhe, Institut f\"ur Kernphysik,\\
D-76021 Karlsruhe, Postfach 3640, Germany\\
e-mail: thomas.jannakos@bk.fzk.de}

\maketitle\abstracts{
KARMEN, the {\bf KA}rlsruhe {\bf R}utherford {\bf M}edium {\bf E}nergy 
{\bf N}eutrino experiment, is located at
the pulsed spallation neutron source ISIS of the Rutherford Appleton 
Laboratory. In the ISIS beam stop \num , \nue\ and \numb\ are produced from
the \pip --\mup\ decay chain at rest with energies up to 52.8 MeV.
Besides a very low \nueb\ contamination, ISIS stands out for its unique
time structure. This allows for a highly sensitive search for 
\Osz --oscillations with the KARMEN detector, a 56\,t segmented liquid
scintillation calorimeter with very good time, energy and position resolution.
In 1996 an additional third veto counter was installed within the 7000\,t
steel blockhouse that shields the detector against cosmic induced background.
Covering the detector from all sides it strongly reduces the background
of cosmic induced high energy neutrons by a factor of 40.
Here we present the data taken after this major upgrade from Feb. 1997 until
Feb. 1999. 
Since there is no indication for any beam excess, an upper limit for the
\Osz\ oscillation is deduced using the Unified Approach based on a maximum 
likelihood analysis. The result, as all the data presented before, questions
the interpretation of the LSND beam excess as an indication for
\Osz\ oscillation.
}

\newpage

\pagenumbering{arabic}
\setcounter{page}{2}
\section{Introduction}
The search for neutrino oscillations is one of the most fascinating topics
of modern particle physics. The {\bf KA}rlsruhe {\bf R}utherford {\bf M}edium 
{\bf E}nergy {\bf N}eutrino experiment KARMEN searches for neutrino 
oscillations in different appearance (
\mbox{\num $\rightarrow\,$\nue } \cite{zeitnitz} and 
\mbox{\numb $\rightarrow\,$\nueb } ) and
disappearance modes (\mbox{\nue $\rightarrow\,$\nux }\cite{nuex}). 
The physics program of KARMEN
also includes the investigation of $\nu$--nucleus interactions \cite{reinhard}
as well as the search for lepton number violating decays of pions and muons
and a test of the V--A structure of the \mup\ decay \cite{omega}.
In the following we present the result of the search for \Osz\ oscillation
on the basis of the data taken 
from February 1997 until February 1999 (KARMEN\,2 data)
after the experiment upgrade in 1996.
The data taken before the upgrade from 1990 - 1995 (KARMEN\,1 data)
is not included in the
analysis. Such a combined analysis would yield a much lower sensitivity due to
the relatively high cosmic induced neutron background of the KARMEN\,1 data.\
In the data set presented here we measure the expected number of background 
events. Therefore we used the Unified Approach \cite{cous} based on a
maximum likelihood analysis to derive a 90\,\% confidence interval.

\section{Neutrino Production and Detection}

The KARMEN experiment utilizes the neutrinos produced by the neutron spallation
source ISIS of the Rutherford Appleton Laboratory in Chilton, Oxon, UK.
An intense beam ($200\,\mu$A) of protons is accelerated to an energy of
800\,MeV by a rapid cycling synchrotron. The two
parabolic proton pulses of 100\,ns base width and a gap of 225\,ns are
produced with a repetition frequency of 50\,Hz (duty cycle is $10^{-5}$).
The protons are stopped in the compact tantalum
beam stop. Apart from spallation neutrons a large number of pions is produced
and stopped immediatly within the target.
While almost all \pim\  undergo nuclear capture, the \pip\ decay at rest (DAR)
into \mup\ and  \num .
The \mup\ are also stopped within the target and decay at rest via
$\mup\rightarrow\pos\,+\,\nue\,+\,\numb$. The minor fraction of \pim\
that decays in flight ($0.65\,\%$ relative to \pip\ DAR) with an again suppressed
subsequent \mum\ decay leads to an extreme small \nueb\ contamination
  of $ \nueb /\nue \, \le \, 6.2\cdot 10^{-4}$ \cite{bob}.
The energy spectra of the neutrinos are well defined due to the DAR
of both the \pip\ and \mup . The \num\ from \pip -decay is monoenergetic 
with E(\num)=29.8\,MeV; the continuous energy distributions up to 52.8\,MeV
of the \nue\ and \numb\  can be calculated
using the V-A theory and show the typical Michel shape.
Therefore ISIS is a unique, isotropic source of \num , \nue\ and \numb\ from 
\pip -\mup\ DAR that stands out for its time structure, the small \nueb\
contamination and the well defined time and energy distribution of the 
produced neutrinos.

These neutrinos are detected with the KARMEN detector, a segmented calorimeter
of 56\,t of liquid scintillator. The matrix structure consists of 512 
($32\,rows\,\times16\,columns$) optically independent modules with a cross
section of $17.4\,\times\,17.8\,cm^2$ and a length of 353\,cm.
The segmentation is made of thin double acrylic walls separated by a small air
gap. Every module is read out by two 3\,inch photo tubes at each end.
The position of an event within one module is given by 
the time difference between the photo tubes at both ends.
The optimized optical properties of the organic liquid scintillator
and an active volume of 96\% result in an energy resolution of
$\sigma_E=11.5\% / \sqrt{E [MeV]}$. Gd$_2$O$_3$
coated paper within the module walls provides an efficient detection of thermal
neutrons owing to the very high capture cross section of the \Gdng\ reaction
($\sigma \approx 49000$\,barn). The KARMEN electronics is synchronized to the
ISIS proton pulses to an accuracy of 2\,ns to fully exploit the time
structure of the neutrinos. The detector is well protected against beam 
correlated background as well as the
hadronic component of the cosmic radiation by a blockhouse made of
7000\,t of steel. Cosmic muons entering or stopping close to the detector are
identified by the two inner veto counters.
The innermost veto covers the calorimeter from four sides and consists of 
modules identical to those of the calorimeter but half their width. The second
veto counter is made of 136 plastic scintillator modules that shield the 
detector from five sides.
With this configuration (KARMEN\,1), the dominant background for the search for 
\Osz\ oscillations were high energetic neutrons produced by cosmic muons
within the steel blockhouse. To eliminate this background source an additional
third veto counter made of 136 plastic szintillator modules with a total
area of 300\,m$^2$ was installed in 1996 \cite{drexlin}.
It was placed right inside the steel blockhouse such that every muon is 
detected that could produce a neutron within the blockhouse at a distance of 
up to 1\,m from the detector. With this configuration (KARMEN\,2) cosmic
induced background is considerably reduced by a factor of 40.

\section{\Osz\ oscillation signature}

The probability for \Osz\ oscillations can be written in a
simplified 2 flavor description as
\vspace{-.5ex}
\be
\rm{P}(\bar{\nu}_{\mu}\rightarrow\bar{\nu}_{e}) = \rm{sin}^2\,(2\,\Theta)
\cdot sin^2(1.27 \frac{\Delta \rm{m}^2 L}{E_{\nu}})
\vspace{-.5ex}
\ee
where L is given in meters, $E_{\nu}$ is the neutrino energy in MeV, 
and \dm\ denotes the
difference of the squared mass eigenvalues $\dm = |m^2_1 - m^2_2|$ in \eVc .
The signature for the detection of a \nueb\ is a spatially correlated,
delayed coincidence of a positron from \Png\ with energies up to
$E_{e^+}=E_{\nueb}-Q=(52.8-1.8)\,\rm{MeV}=51.0$\,MeV followed by the $\gamma$ emission 
of either of the two neutron capture processes \Pn\ or \Gn .
The \Pn\ reaction leads to one $\gamma$ with an energy of 
$E(\gamma)=2.2$\,MeV whereas the \Gn\ process leads to 3 $\gamma$ in 
average with a sum energy of 8\,MeV.
The positrons are expected with a 2.2\,$\mu$s\ exponential decrease of 
to the \mup\ decay after beam on target. The time difference between the 
positron and the capture $\gamma$ is given by the thermalization and 
diffusion of the neutron.
\newpage
\begin{figure}[thb]
\centerline{\psfig{figure=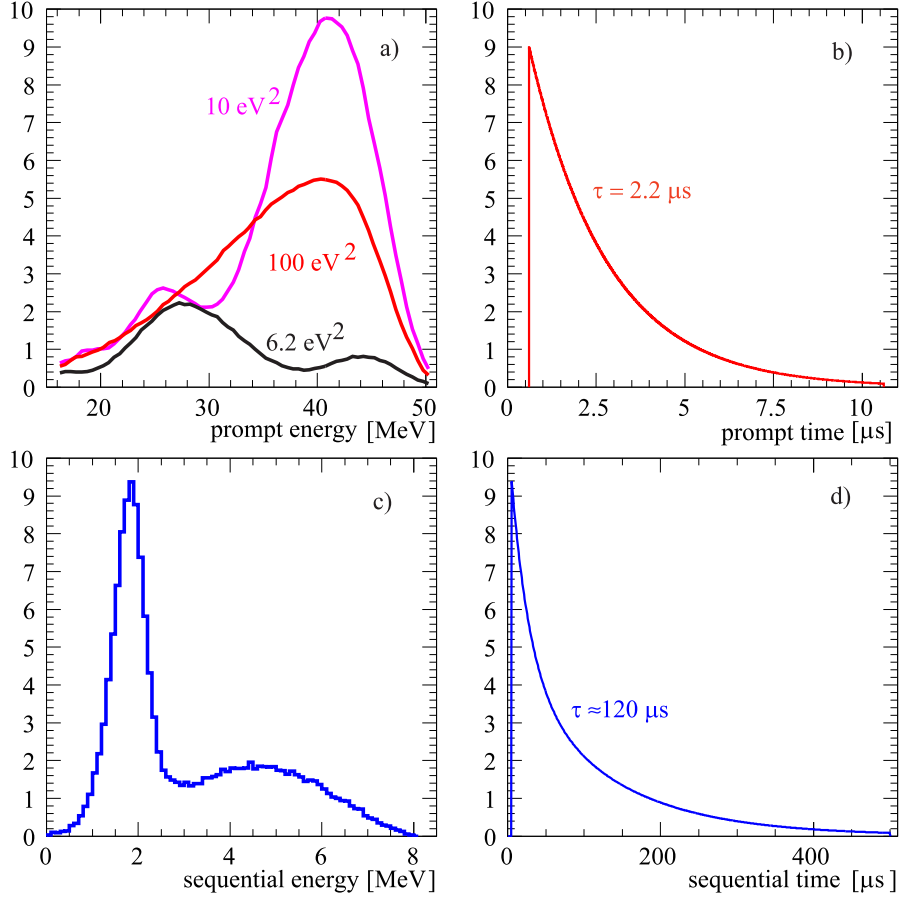,height=12.0cm}}
\caption{Signature of sequences of a positron (prompt event) and the
 correlated gammas from the neutron capture reaction (sequential event) that
 are expected for the \Osz\ oscillation in the KARMEN detector:
 a) visible energy of the positron for three different values of \dm\
 as given by a Monte Carlo (MC) simulation; b) time of the positron
 relative to beam on target; c) visible energy of the delayed gammas 
 from the nuclear neutron capture on either the free protons in the scintillator 
 or the gadolinium in the  segmentation; d) time difference of the neutron
 capture reaction relative to the positron.
\label{fig:signat}}
\end{figure}
To suppress cosmic induced background, a positron candidate is accepted 
only if there is no activity in the central detector and in both inner veto
counters up to 24\,$\mu$s\ before. When only the outermost third veto counter 
was hit, a dead time of 14\,$\mu$s\ is applied.\\
The unique signature of the \Png\ reaction allows already for a strong
discrimination of cosmic and neutrino induced background.
The following cuts are introduced to maximize the sensitivity of the
experiment: The positron has to be detected in a time window from
$0.6-10.6\,\mu$s after beam on target with its energy in the range
from $16-50$\,MeV. The sequential gamma must have an energy below 8 MeV
and has to be correlated in space (within 1.2\,m$^3$)
and time ($5-300\,\mu$s) to the positron. For these cuts the total 
detection efficiency is -- slightly depending on \dm\ -- 
approximately 20\,\%. The expected signature for oscillation sequences
in the KARMEN detector is shown in Fig. 1.

\begin{table}[h]
\caption{Expected sequences from background reactions within the cuts
 specified above. Given are the mean values and their errors. Shown in the 
 two last  rows are the number of expected \Png\ reactions from the \Osz\ 
 oscillation for  high \dm\ ($\,=\,100\,eV^2$) assuming maximal mixing 
 (i.e. $\sn\,=\,1$) and the number of actually measured sequences.
 All numbers are given for two different energy windows from $16-50$\,MeV
 and $36-50$\,MeV respectively.
 \label{tab:exp}}
\begin{center}
\begin{tabular}{|l|l|l|}
\hline
 Background reaction  	& events {\small (E$\ge$16MeV)}
& events {\small(E$\ge$36MeV)}\\ \hline
 \Cng\ reaction			& $2.6\,\pm\,0.3$& $0.00\,\pm\,0.01$ \\
 $\nu$ induced random coincidences& $2.3\,\pm\,0.3$& $0.09\,\pm\,0.03$ \\
 \nueb\ contamination from ISIS	& $1.1\,\pm\,0.1$& $0.31\,\pm\,0.03$ \\
 cosmic induced background	& $1.9\,\pm\,0.1$& $0.56\,\pm\,0.07$ \\ \hline
 total expected background	& $7.8\,\pm\,0.5$& $0.97\,\pm\,0.08$ \\ \hline\hline
 measured sequences		&    8           &   0               \\ \hline
 \Png\ reactions for \sn$\,=1$	& $1605\,\pm\,176$& $712\,\pm\,78$ \\ \hline
\end{tabular}
\end{center}
\end{table}

\section{Background Sources}

One of the main advantages for the search of \Osz\ oscillations with the
KARMEN experiment is that the expected background is not only very small
but also known with a high percision because most of it can be independently
measured by applying different cuts. There are only four different sources 
of background:
\begin{itemize}
\setlength{\parskip}{0.0ex}
\setlength{\partopsep}{0.0pt}
\setlength{\parsep}{0.0ex}
\setlength{\itemsep}{0.0ex}
\setlength{\topsep}{-1.5ex}
\item \nue\ induced sequences caused by the charged current reaction
      \Cng\ where the subsequent $\beta$ decay of the \Nzg\ ($\tau = 15.9$\,ms)
      occurs within the first 300\,$\mu$s.
\item Neutrino reactions that have a random coincidence with a low energy event
      from the natural radioactivity inside the detector.
\item The small intrinsic \nueb\ contamination from the \pim -\mum\ decay chain in
      the ISIS target.
\item Undetected cosmic muons which enter the detector or produce high energy
      neutrons via deep inelastic scattering in the inner part of the steel
      blockhouse.
\end{itemize}
The only background source not accessible to direct measurement is the 
\nueb\ contamination. It is calculated using a detailed MC simulation of the 
ISIS target as well as all pion and muon production and decay or capture 
reactions \cite{bob}.
Table 1 lists all background reactions and gives the number of expected events 
as well as their errors for the above defined cuts.

\section{Maximum Likelihood Analysis}

The maximum likelihood (ML) analysis is the most powerful method to infer the 
strength of a possible signal or to derive an upper limit if such a signal
is  not seen.
Because of some advantages over other methods we use here the 
Unified Approach \cite{cous} recommended by the PDG \cite{pdg} to derive 
a 90\,\% confidence interval from our ML analysis.
For this ML analysis every background reaction and a possible oscillation 
signal are taken into account with their different probability density
functions for the time and energy of the prompt event (the \pos )
as well as the energy, and the time and position difference of the 
sequential event (the neutron capture) relative to the prompt event.
The relative contributions of the individual background sources to the total
number of background sequences is fixed whereas the number of 
oscillation sequences is allowed to vary freely. 
As an additional information, the likelihood function 
(LF) is weighted with a factor that is the conditional poisson probability 
of the number of inferred background sequences given the expectation value of 
the total background. The resulting LF depends on \dm\ and \sn\ only, and thus 
for a given \dm\ only on 
the number of oscillation sequences N$_O$ infered (or the number of background
sequences N$_B$ , for N$_B$=N$_{total}$-N$_O$).

For the Unified Approach we divided the relevant [\dm ;\sn ] parameter space
in the interval [$(10^{-2}eV^2,10^{2}eV^2);(10^{-4},1)$] using a 
logarithmically equidistant grid of $90\times72$ points \cite{mark}.
At every point on the grid we generate 8000 MC data samples according
to the expected background and the given values of \dm\ and \sn\ for this point.
To these data samples the same ML analysis as to the experimental sample
is applied. For every MC sample of this specific point on the grid one
calculates the logarithm of the likelihoodratio \lmc\ of the value of the LF
at its global maximum in the [\dm ;\sn ] parameter space to the value of
the LF at the given point on the grid for which the sample was generated.
This procedure gives a characteristic MC generated distribution of \lmc\ for
every point on the grid which is then compared to the \lex\ value of the 
experimental data set (i.e. the logarithm of the ratio of the experimental LF
at its global maximum to its value at a given point on the grid).
The 90\,\% confidence interval (C.I.) is the set of points on the grid for 
which the experimental value \lex\ is smaller than at least 10\,\%
of all MC generated \lmc\ (see Fig. 2). If, on the other hand, 
for given parameters
[\dm ;\sn ] \lex\ lies in the upper 10\,\% tail of the MC distribution
this point on the grid does not belong to the 90\,\% confidence interval.
The upper 90\,\% confidence limit (C.L.) as shown in Fig.\,4 is the 
upper limit of the 90\,\% C.I.. The interpretation of this C.L. is that 
for all parameter combinations \dm\ and \sn\ on this curve 90\,\% of
a large number of hypothetical KARMEN experiments would have seen a larger 
``signal'' (i.e. {\it smaller} \lex ) than the one actually observed
if -- and this is important -- the true parameters of the \Osz\ oscillation
were in this region of the parameter space.
\begin{figure}[htb]
\centerline{\psfig{figure=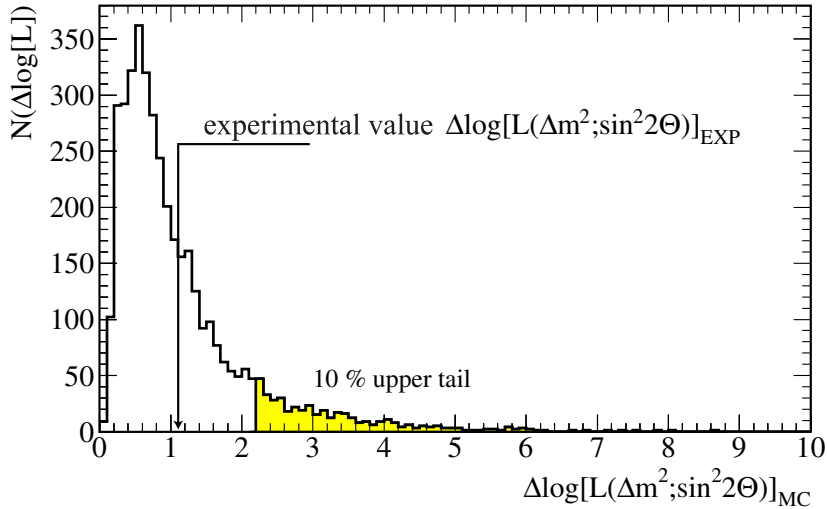,height=7.0cm}}
\caption{Monte Carlo  generated distribution of the logarithm of the 
 likelihoodratios
 \lmc\ for \dm$=100$\,eV$^2$ and \sn$=0.001$. The experimental value \lex\
 for this point in the parameter space is to the left of the upper 10\,\%
 tail of the MC distribution. Therefore this parameter combination belongs
 to the 90\,\% C.I..
\label{fig:signat}}
\end{figure}

\vspace{-2ex}
\section{Results and Conclusion}

The results presented here are based on the data recorded in the
measuring period from February 1997 to February 1999 which corresponds to
4670\,C protons on target.
Within the cuts defined in Sect.\,3 we find 8 sequences as shown in Fig.~3.
Since we expect a total background of $7.8\,\pm\,0.5$ sequences there is
absolutely no indication for a beam excess.
\begin{figure}[htb]
\centerline{\psfig{figure=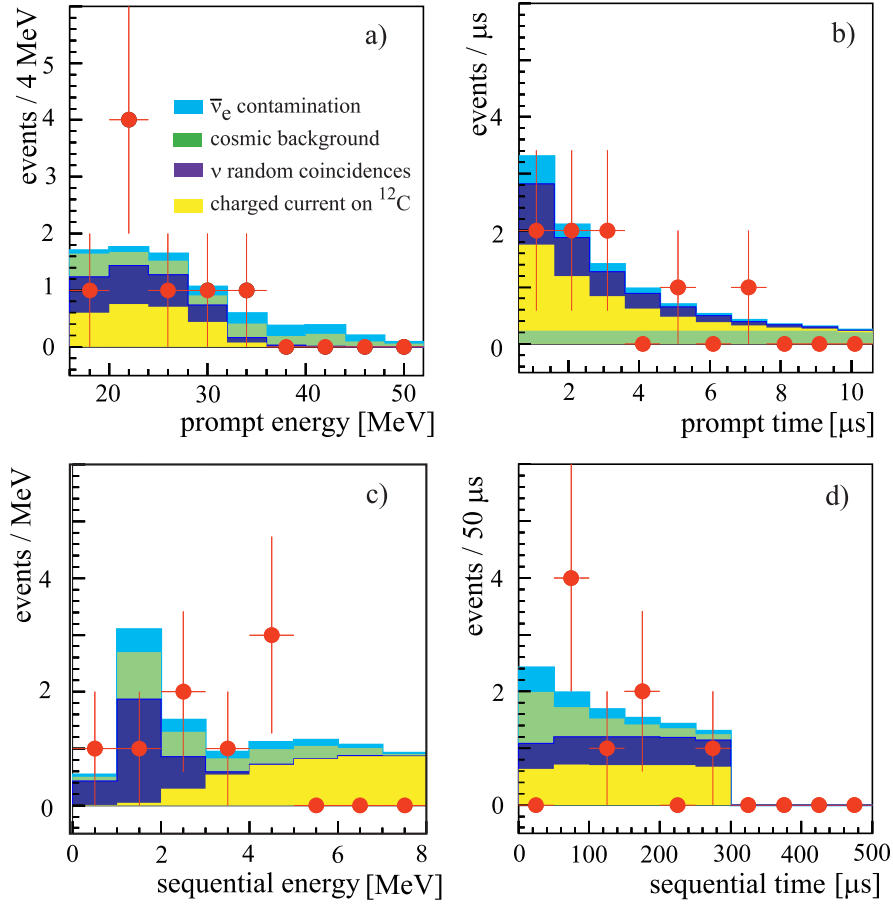,height=12.0cm}}
\caption{Distribution of the expected background sequences for the prompt
 energy (a) and time (b) distribution as well as the sequential
 energy (c) and time (d) distribution. Also shown are the 8 measured
 sequences which agree nicely in their shape with the expected background.
\label{fig:signat}}
\end{figure}
For this data set the above described analysis leads to a 90\,\% C.L. 
of \sn=$2.1\cdot10^{-3}$ for large \dm\ (i.e. 100\,eV$^2$). 
The 90\,\% C.L. as a function of \dm\ can be seen in Fig.4. 
Also shown is the sensitivity of the KARMEN experiment. The sensitivity of 
an experiment is defined as the mean confidence limit a large number of
identical experiments would yield if there was no oscillation.
The actual limit is slightly ``better'' then the sensitivity with 
\sn=$2.3\cdot10^{-3}$ for large \dm .
Na$\ddot{\rm{\i}}$vely one would expect that the sensitivity is at
slightly lower \sn\ than the exclusion curve and not vice versa 
because we measure more events (0.2) than the expected background. 
This apparent contradiction is explained by the fact that there are no 
events above 36 MeV where
-- depending on \dm\ -- roughly half of all oscillation sequences should be 
and only $0.97\,\pm\,0.08$ background events are expected (see Tab.1). 
This leads also to an LF that has its global maximum in the region of
small negative \sn\ which - for a null result - is as possible as a global 
maximum in the region of positive \sn .
The 90\,\% C.L. from our ML analysis is compared in Fig.\,4 to the LSND result 
\cite{lsnd}. It excludes most of the LSND favoured region and is  thus
-- as is the fact that there is no event above 36\,MeV -- strongly
questioning the interpretation that the LSND beam excess is an indication
for \Osz\ oscillations.
Furthermore one has to keep in mind that the KARMEN limit that was derived
from the Feb.97 - Apr.98 data set and that is also shown in 
Fig.\,4 puts an even stronger constraint on the LSND favoured region.
In this data set (which had slightly different cuts) we expected 
$2.9\,\pm\,0.1$ background sequences and measured no event at all.
This yields, using again the unified approach, a 90\,\% C.L. of 
$1.3\cdot10^{-3}$ for large \dm\ and a sensitivity
of $5.4\cdot10^{-3}$, respectively \cite{neut98}, \cite{thom}. Although 
this limit was
criticized by some people, it is of course valid and correct:
If one accepts the ansatz of the Unified Approach one must NOT argue that 
this (``lucky'') limit is not trustworthy for this would 
mean to reduce the Unified Approach to absurdity.

\begin{figure}[htb]
\centerline{\psfig{figure=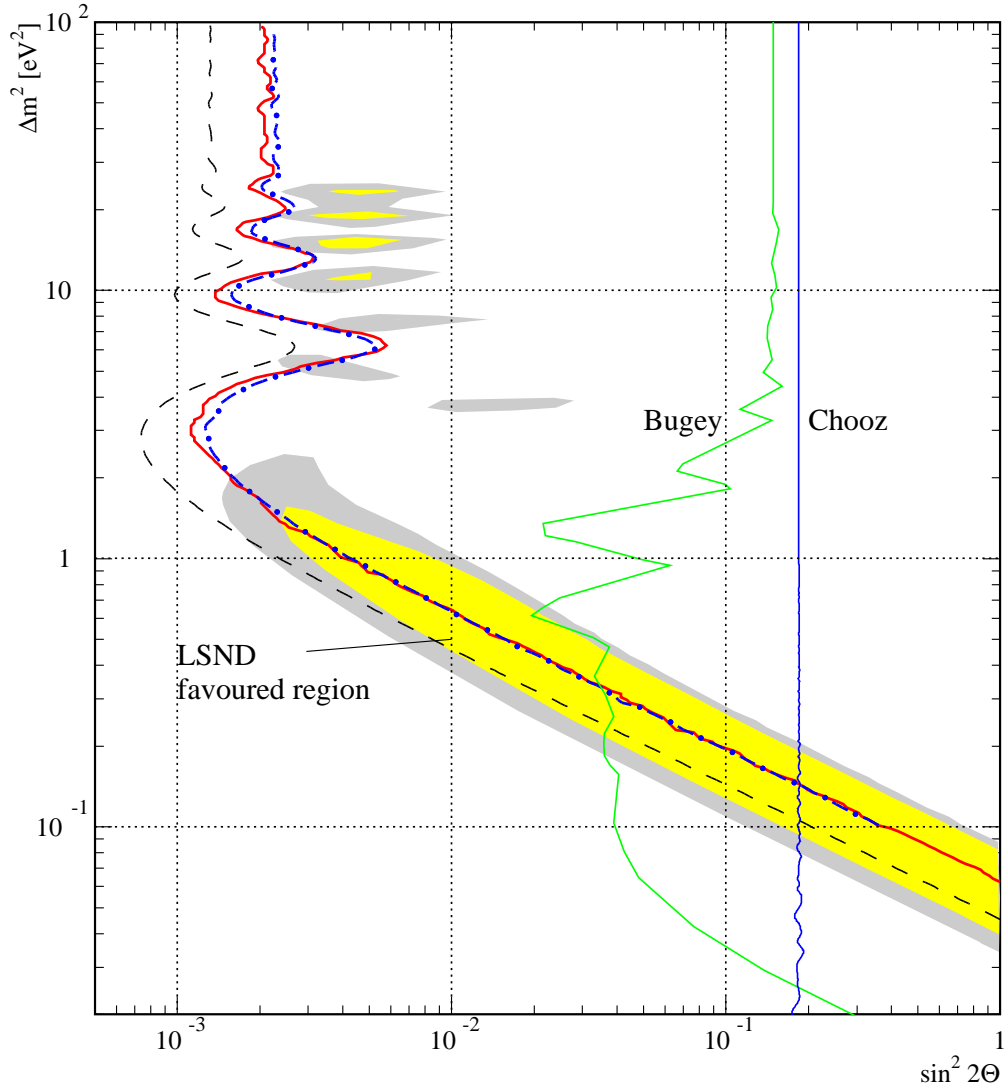,height=14.0cm}}
\caption{KARMEN\,2 90\,\% confidence limits according to
	the Unified Approach compared to other experiments: The full line
        is the 90\,\%\,C.L. of the data presented here, the dotted line the 
	corresponding sensitivity and the dashed line the 90\,\%\,C.L. 
	derived from the Feb.\,97-Apr.\,98 data. Also shown are the 
	90\,\%\,C.L. of the two reactor experiments Chooz \cite{chooz} and 
	Bugey \cite{bugey} and the favoured region for \Osz oscillations as
	reported by the LSND experiment\cite{lsnd}. Areas to the right of
	the 90\,\% C.L. are excluded with a probability of more than 90\,\%.
	For the LSND result the ``99\,\% favoured region'' (total shaded 
	area) and the ``90\,\% favoured region'' (light-shaded area)
	are given.\label{fig:limit}}
\end{figure}

\end{document}